# Engineering Canalized Near-Field Radiative Heat Transfer between Fixed Nanoparticles with Twisted α-$MoO_3$ Trilayers

Yang Hu[1,*], Pablo Alonso-González[1,2,*]

[1]Department of Physics, University of Oviedo, Oviedo 33006, Spain

[2]Center of Research on Nanomaterials and Nanotechnology, CINN (CSIC-Universidad de Oviedo), El Entrego 33940, Spain

[*]Corresponding authors: Yang Hu (hu_yang@mail.nwpu.edu.cn), Pablo Alonso-González (pabloalonso@uniovi.es)

## Abstract

Canalized phonon polaritons in twisted van der Waals crystals offer a route to control nanoscale thermal energy flow. Building on the established twisted α-$MoO_3$ trilayer platform, we examine how internal twist angles modify near-field radiative heat transfer (NFRHT) between a prescribed nanoparticle pair while both the nanoparticle axis and the bottom-layer crystallographic orientation remain fixed. Under this constraint and at equal total α-$MoO_3$ thickness, the optimized trilayer reaches 2.84 times the bilayer maximum radiative power, while its mean power over the sampled internal twist configurations is 1.51 times the bilayer value. The second internal twist coordinate also enlarges the set of configurations supporting strong transfer, with local angular tolerance when either internal angle is varied separately. Momentum-space Green-tensor diagnostics and real-space energy-density maps associate the enhancement with stronger receiver-directed scattering and improved channel alignment. Selected trilayer configurations retain enhanced coupling across the sampled family of nanoparticle resonances and at long distances. These results quantify the additional internal configurational freedom available for engineering radiative heat transfer under a fixed laboratory-frame geometry.

Multilayer van der Waals (vdW) materials have emerged as a versatile platform for controlling nanoscale light–matter interactions[1-7]. Their weak interlayer bonding allows individual layers to be stacked with arbitrary relative orientations, introducing twist as an additional degree of freedom beyond composition, thickness, and geometry [4, 8, 9]. Apart from its prominent role in electronic systems, where it has enabled phenomena such as flat-band superconductivity, moiré excitons, and interlayer magnetism, twist has also become a powerful concept in nanophotonics and polaritonics[7, 10-18]. In polar vdW crystals, in particular, interlayer rotation provides a direct means to reshape the dispersion of phonon polaritons and thus to control the direction of the energy flow at the nanoscale[19-22]. Among these materials, $\alpha$-phase molybdenum trioxide ($\alpha$-$MoO_3$) is especially attractive because of its strong in-plane anisotropy and natural mid-infrared hyperbolic response[23, 24]. In twisted $\alpha$-$MoO_3$ bilayers, the coupling between rotated anisotropic layers redesigns the polaritonic isofrequency contours, driving a transition from open hyperbolic-like curves to closed elliptic-like contours[1, 2, 7, 25, 26]. At a critical twist angle, known as the photonic magic angle, the isofrequency contour becomes nearly flat, giving rise to highly directional and nearly diffractionless polariton canalization[2, 7]. This topological transition provides an effective mechanism for steering nanoscale electromagnetic energy without relying on external fields or complex nanopatterning [2, 7, 25, 26], in contrast to platforms that require electrical tuning or patterned plasmonic elements [8, 27]. Related advances in complex-frequency sensing, hot-carrier dynamics, ultrafast nano-imaging, and van der Waals quantum-light sources further illustrate the breadth of nanoscale light–matter interactions [28-31]. Recent analytical work has extended twistoptics to planar heterostructures containing an arbitrary number of rotated finite-thickness anisotropic layers and two-dimensional conductive sheets [32].

This type of directional polariton propagation is directly relevant to near-field radiative heat transfer (NFRHT), where evanescent waves and polaritonic modes dominate thermal energy transfer at subwavelength separations[33-45]. In planar slab-slab geometries, the heat flux is determined by integrating the energy transmission coefficient over the entire momentum space, so the overall polaritonic density of states plays a central role[19, 20, 46]. Under these conditions, closed elliptic dispersions can provide larger integrated contributions than open hyperbolic ones[20]. However, the underlying mechanism is fundamentally different for heat transfer between two nanoparticles mediated by a polaritonic slab[34, 47-52]. In this case, the exchange is governed by the dyadic Green's function connecting the thermally fluctuating dipoles, and the relevant enhancement depends not only on the polaritonic density of states, but also on how their spatial distribution and energy-flow direction match the relative position of the nanoparticles[22, 48, 49].

Environmental control of nanoparticle NFRHT includes multibody thermal switching [53], particles near spherical objects [54], spherical or parallel-surface confinement [55,56], and many-particle exchange [57,58]. Building on this broader framework, we examine how internal rotations in a biaxial multilayer modify the radiative coupling between prescribed nanoparticle positions.

Surface-mediated nanoparticle NFRHT has previously been demonstrated through propagating surface phonon polaritons on SiC [47], electrically tunable anisotropic surface plasmon polaritons in patterned graphene-strip metasurfaces [34], and high-wavevector modes in hyperbolic multilayers [59]. Twisted graphene gratings provide an additional route for controlling directional thermal transport [22]. Previous studies have shown that canalized polariton modes in twisted α-$MoO_3$ bilayers can be used for configurable thermal routing between spatially separated terminals [60]. Here, the emphasis is instead on a prescribed emitter–receiver pair, for which the additional internal twist coordinate of the trilayer enlarges the set of channel-matched configurations while keeping both the nanoparticle axis and the bottom-layer crystallographic frame fixed.

Twisted trilayer hyperbolic-polaritonic structures have previously been explored as a route to broadband dispersion engineering through interlayer evanescent coupling, including applications to tunable NFRHT [61]. In the present work, α-$MoO_3$ is selected because its naturally biaxial response supports mid-infrared hyperbolic phonon polaritons without requiring lateral patterning or electrostatic tuning [23,24]. Its strong in-plane anisotropy and comparatively low phonon damping favor highly confined and directional polariton propagation, while experimentally demonstrated α-$MoO_3$ trilayers provide multiple spectrally robust canalization conditions [7]. Together, these properties make α-$MoO_3$ particularly well suited for engineering long-range polaritonic channels between prescribed nanoparticle positions.

We employ the experimentally established twisted α-$MoO_3$ trilayer [7] to investigate radiative heat transfer between a prescribed nanoparticle pair. When both the nanoparticle axis and the bottom-layer crystallographic orientation are fixed, the bilayer provides one internal twist variable whereas the trilayer provides two. The additional internal coordinate enlarges the set of configurations capable of producing strong receiver-directed coupling without globally rotating the support or repositioning the nanoparticles.

We begin by considering NFRHT between two nanoparticles mediated by a twisted α-$MoO_3$ bilayer. The geometry is illustrated in **Figs. 1**(a) and (b). Two nanoparticles of radius *R* are placed above the bilayer slab, separated by an interparticle distance *d* and located at a distance $h$ from the slab surface. The bottom α-$MoO_3$ layer is kept fixed, with its [100], [001], and [010] crystallographic directions aligned with the $x$-, $y$-, and $z$-axes, respectively. The top layer is rotated in the $x$-$y$ plane by an interlayer twist angle $\beta_s$. In addition, the relative orientation of the nanoparticle pair can be varied by rotating the receiver nanoparticle around the emitter in the $x$-$y$ plane. The angle between the interparticle axis and the $x$-axis is denoted by $\beta_p$. The dielectric response of the nanoparticles is described by a Drude model,

$$\varepsilon(\omega) = 1 - \frac{\omega_p^2}{\omega^2 + i\Gamma\omega}, \tag{1}$$

where $\omega_p = 1588.3\text{cm}^{-1}$ and $\Gamma = 0.001\omega_p$. With these parameters, the dipolar resonance of the nanoparticles occurs at approximately 917 $\text{cm}^{-1}$, within the Reststrahlen band of α-$MoO_3$.

The nanoparticles are modeled as generic isotropic Drude spheres and are not fitted to a specific material. The selected plasma frequency places the weak-damping quasistatic dipolar resonance at approximately 917 $\text{cm}^{-1}$, thereby isolating spectral matching with the α-$MoO_3$ phonon-polariton response. Experimental implementations require a sufficiently low-loss localized resonance overlapping the relevant Reststrahlen band, for which tunable doped-semiconductor or conducting-oxide nanocrystals provide possible realizations [62, 63]. Polar dielectric nanoparticles such as SiC offer an alternative through localized surface phonon-polariton resonances and should be described by an appropriate Lorentz dielectric response [22]. The damping $\Gamma$=0.001$\omega_p$ is an idealized low-loss choice for the present mechanism study.

The nanoparticles are described by a generic isotropic Drude model. Each α-$MoO_3$ flake is instead described by a biaxial Lorentz tensor with principal axes $x$, $y$, and $z$ along [100], [001], and [010], respectively. The in-plane twist is introduced by rotating this tensor for each flake. All flakes are laterally infinite in the *x-y* plane and form an ideal direct-contact stack. The finite rectangles in the schematics indicate crystallographic orientation only; both nanoparticles remain above the common overlap region of all layers. Vacuum occupies both exterior half-spaces; no supporting substrate, positioning tip, spacer, or trapping structure is included in the Green tensor. Unless otherwise noted, $R$=10nm, $d$=1000nm, $h$=150nm, $T_1$=350K, and $T_2$=$T_s$=300K. In the principal bilayer–trilayer comparison, the total α-$MoO_3$ thickness is fixed at 600 nm: 2×300 nm for the bilayer and 3×200 nm for the trilayer. The Lorentz parameters are taken from Ref. [24] and listed in the Supporting Information.

The NFRHT is calculated within fluctuational electrodynamics and the electric-dipole approximation. The reported $P$ is the net radiative power transferred from emitter nanoparticle 1 to receiver nanoparticle 2, rather than the total power absorbed by particle 2 from all surrounding bodies. The α-$MoO_3$ slab and particle 2 are maintained at the same temperature, $T_s$=$T_2$=300 K, so their mutual net thermal exchange vanishes by detailed balance. The slab nevertheless remains fully included electromagnetically through the scattering Green tensor, which describes its modification of the particle-particle coupling. The radiative power between the nanoparticles can be written as [49, 64, 65]:

$$P_i = \int_0^\infty \frac{d\omega}{2\pi} P_{i,\omega} = 3\int_0^\infty \frac{d\omega}{2\pi} [\Theta(T_j) - \Theta(T_i)]\tau_{ji}, \quad (2)$$

where Θ(ω,T) is the mean Planck oscillator energy and $\tau_{ji}$ is the transmission coefficient from particle j to particle i; j=1 and i=2 for the reported emitter-to-receiver transfer. For the identical isotropic particles considered here, the transmission coefficient is evaluated from the complete dyadic Green tensor, so all nine source–receiver components and all three uncorrelated thermal-dipole orientations are included. The angle $\beta_p$ denotes the azimuth of the interparticle axis, not a prescribed dipole polarization. The dipole approximation is well satisfied over the considered parameter range. At 917 cm$^{-1}$, $k_0R \approx 5.8 \times 10^{-3}$; even at the smallest center-to-center distance in Fig. 4, $R/d = 0.10$, while $R/h \approx 0.067$. Higher-order multipoles are therefore expected to remain negligible for the particle sizes and separations considered here. The Green tensor contains vacuum and scattering contributions, with the latter describing the α-$MoO_3$-mediated coupling; full expressions are provided in the Supporting Information.

Related nanoscale radiative thermal-switching experiments based on multibody effects are reported in Ref. [53]. The present study focuses on the channel-matching mechanism in an idealized geometry rather than a device-specific implementation.

Figure 1(c) shows the bilayer orientation-and-twist map $P(\beta_p,\beta_s)$, with $\beta_s$ on the horizontal axis and $\beta_p$ on the vertical axis. Because $\beta_p$ rotates the particle axis relative to the fixed bottom-layer frame, this two-coordinate map is a global reference over the two displayed coordinates rather than a strict fixed-endpoint tolerance map. Its maximum is $1.87\times10^{-14}$ W at $(\beta_p,\beta_s)$=(44°,106°), whereas the minimum is $7.72\times10^{-17}$ W at $(\beta_p,\beta_s)$=(90°,0°), giving a modulation ratio of approximately 243. For the strict fixed-endpoint benchmark used below, $\beta_p$ is fixed at 0° and only $\beta_s$ is varied. The bilayer then reaches $P_{\mathrm{max,bi}}$=$7.54\times10^{-15}$ W at $\beta_s$=36°, and its mean radiative power over all sampled $\beta_s$ values is $3.53\times10^{-15}$ W. The complete fixed-endpoint statistics are reported in Table S1.

To identify the relevant slab-mediated channels, Figs. 1(*d*–f) and 2(*d*–f) display the *zz* component of the scattering Green tensor in the Weyl representation before integration over $k_x$ and $k_y$. This diagnostic highlights high-k p-polarized phonon-polariton branches but is not used alone to calculate $P$, which contains the complete dyadic tensor.

In Figs. 1(d–f), the green curves show the bilayer isofrequency contours. The emitter–receiver directions are indicated by the black dashed lines in Figs. 1(g–i). The group velocity is normal to the isofrequency contour, so nearly flat contour segments support collimated propagation [1,7]. Efficient transfer also requires sufficient coupling to the receiver.

To quantify receiver-directed propagation, at 917 $cm^{-1}$ we evaluate the scattering and vacuum angular intensities $I_{sc}(\varphi)$ and $I_0(\varphi)$ on a circle centered at the emitter with radius equal to the interparticle separation. From $I_{sc}(\varphi)$, we define $D_2$ as the second angular-moment magnitude, which quantifies the concentration of the scattering response along a preferred axis, $C_{NP} = I_{sc}(\beta_p)/\langle I_{sc}(\varphi)\rangle_\varphi$ as the receiver-directed concentration relative to the angular average, and $\Delta\varphi$ as the angular mismatch between the principal scattering-channel axis and the emitter–receiver direction. For the optimal configurations obtained when both the nanoparticle axis and the bottom-layer crystallographic frame are fixed, $I_{sc}(\beta_p)/I_0(\beta_p)$ is approximately 110 for the bilayer and 309 for the trilayer, while $\Delta\varphi$ decreases from about 8.7° to nearly 0°. These results show that the enhanced radiative transfer under a fixed laboratory frame is associated with both stronger receiver-directed scattering and improved channel alignment. The complete frequency-integrated power is evaluated using $\boldsymbol{G} = \mathbf{G}_0 + \mathbf{G}_{sc}$ and therefore also contains direct coupling and vacuum–scattering interference. The full definitions and numerical diagnostics are provided in Section 4 of the Supporting Information.

For the untwisted case, $\beta_s$=0°, the α-$MoO_3$ bilayer supports a characteristic hyperbolic polariton dispersion. When the two nanoparticles are aligned along the $x$-direction, $\beta_p$=0°, the dominant polariton energy-flow directions are not efficiently matched with the interparticle axis. As a result, the slab-mediated contribution of the hyperbolic modes to the radiative heat transfer enhancement remains limited. This changes when the nanoparticle pair is rotated while keeping the slab untwisted. For $\beta_p$=20° and $\beta_s$=0°, shown in Fig. 1(d), the interparticle axis becomes nearly perpendicular to the hyperbolic isofrequency contours and is therefore approximately aligned with the polariton group velocity. In this configuration, the hyperbolic polariton modes provide a more efficient channel for energy transfer between the two nanoparticles, leading to enhanced radiative power.

For $\beta_p$=0° and $\beta_s$=90°, shown in Fig. 1(e), the isofrequency contour is approximately square-like. Its nearly flat sides support collimated propagation along more than one in-plane direction, rather than the single canalization axis associated with a nearly flat contour at a photonic magic angle [1]. Although part of this response is aligned with the interparticle axis, the radiative power remains below the maximum in Fig. 1(c). Directional alignment alone therefore does not determine the transfer strength.

The strongest enhancement is obtained when both the nanoparticle orientation and the interlayer twist angle are tuned simultaneously. As shown in **Fig. 1**(f), at $\beta_p$=44° and $\beta_s$=106°, the polariton response develops a pronounced canalized distribution that spans an extended region of momentum space and is strongly aligned with the interparticle axis. This favorable directional matching opens efficient slab-mediated energy-transfer channels between the two nanoparticles, resulting in the maximum radiative heat transfer shown in **Fig. 1(**c). These results show that a twisted α-$MoO_3$ bilayer can enhance the NFRHT between nanoparticles by engineering canalized phonon polaritons. For this bilayer, the global maximum at 917 $cm^{-1}$ is obtained by tuning both $\beta_p$ and $\beta_s$. When the particle axis and bottom-layer frame are prescribed, only the internal twist remains available to achieve directional matching.

If global rotation of the complete bilayer relative to the particle pair is allowed, its canalization direction can also be aligned with a prescribed receiver. Even under this global optimization, the trilayer reaches a higher maximum radiative power than the bilayer. The advantage becomes substantially more pronounced when both the nanoparticle axis and the bottom-layer crystallographic frame are fixed. Under this constraint, the bilayer provides one internal twist variable, whereas the trilayer provides two, giving additional configurational freedom without rotating the support or repositioning the nanoparticles.

To complement the momentum-space analysis, Figs. 1(g–i) and 2(g–i) show the emitter-generated spectral electric-field energy density at 917 $cm^{-1}$ in a plane 150 nm above the upper surface, calculated following Yang et al. [22]. This source-resolved quantity provides a direct real-space visualization of the field launched by the emitter and reveals how the resulting polaritonic channel aligns with the prescribed emitter–receiver direction. The maps include all three thermally fluctuating emitter-dipole orientations and all Cartesian electric-field components.

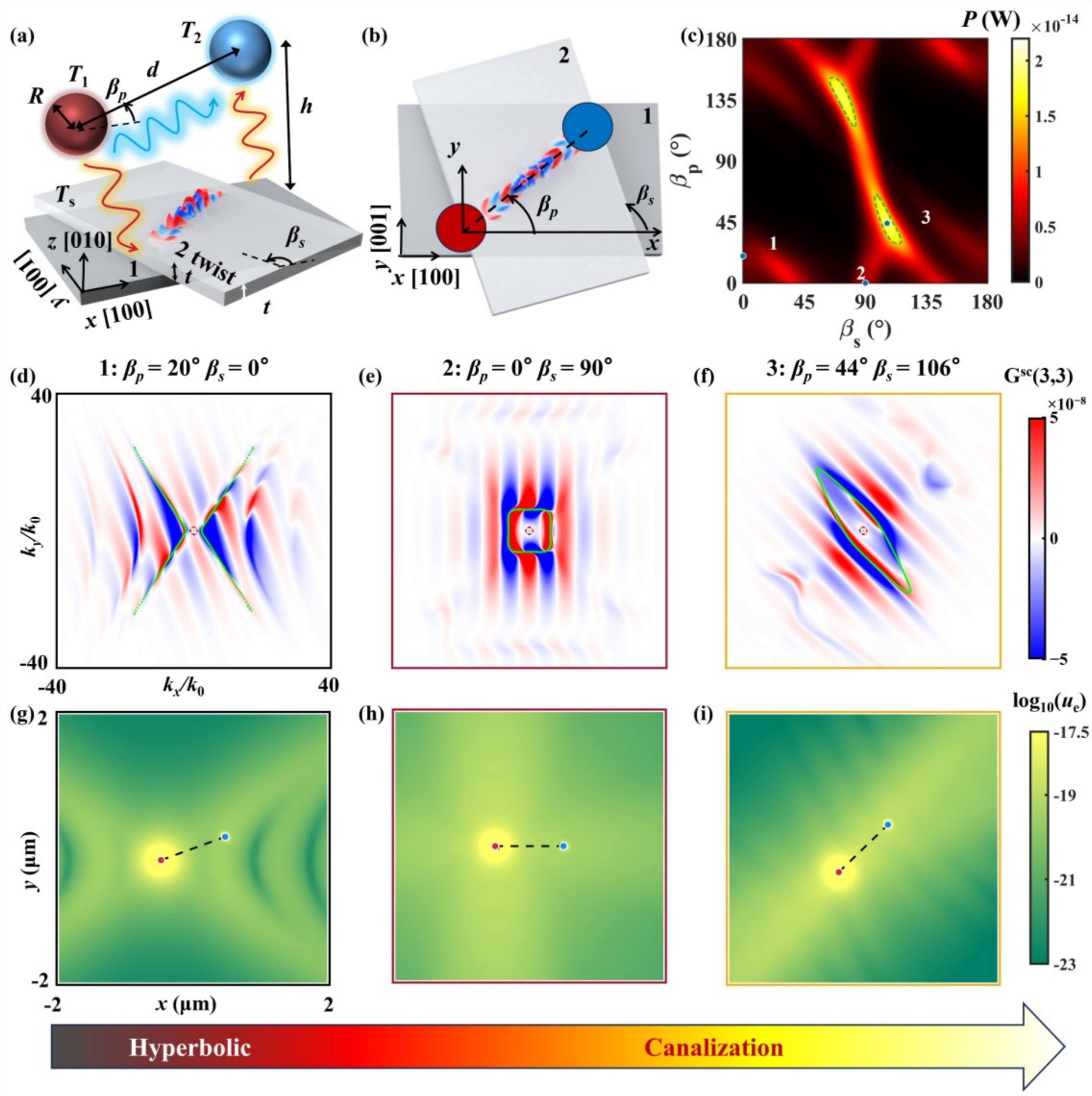

**Fig. 1.** NFRHT between two nanoparticles mediated by a twisted α-$MoO_3$ bilayer. (a) Schematic side view of two nanoparticles placed above a twisted α-$MoO_3$ bilayer. The emitter and receiver nanoparticles have a radius of $R$=10 nm and are separated by an interparticle distance $d$=1000 nm. The nanoparticle-bilayer distance is $h$=150 nm, and the thickness of each α-$MoO_3$ layer is $t$=300 nm. The temperatures of the emitter nanoparticle, receiver nanoparticle, and bilayer are $T_1$=350 K, $T_2$=300 K, and $T_s$=300 K, respectively. The bottom layer is kept fixed, while the top layer is twisted by an interlayer twist angle $\beta_s$. The angle between the interparticle axis and the $x$-axis is denoted by $\beta_p$. (b) Top-view schematic of the twisted bilayer and the nanoparticle arrangement. The layers are laterally infinite; their finite outlines indicate crystallographic orientation only. (c) Net radiative power $P$ between the two nanoparticles as a function of the interlayer twist angle $\beta_s$ and the nanoparticle orientation angle $\beta_p$. Numbered points identify the three configurations shown below, and green dashed outlines highlight high-power regions. (d–f) Momentum-space distributions of the scattering Green's function component $\mathrm{Re}\left[\mathbf{G}_{\mathrm{zz,sc}}\left(k_x, k_y\right)\right]$ at 917 cm$^{-1}$ for representative angle combinations: (d) $\beta_p$=20°, $\beta_s$=0°; (e) $\beta_p$=0°, $\beta_s$=90°; and (f) $\beta_p$=44°, $\beta_s$=106°. The green curves indicate the isofrequency contours of the twisted α-$MoO_3$ bilayer. (g–i) Corresponding emitter-generated spectral electric-field energy-density maps at 917 cm$^{-1}$ in a plane 150 nm above the upper surface. Red and blue circles mark the emitter and receiver, while black dashed lines indicate their connecting direction. The bottom arrow schematically indicates the evolution from hyperbolic polariton propagation to canalized phonon-polariton transport.

We next use a twisted α-$MoO_3$ trilayer to test whether a second internal variable broadens the set of channel-matched states. Layer 1 and the nanoparticle pair are fixed along the laboratory *x*-axis, while layers 2 and 3 are oriented relative to layer 1 by $\beta_{12}$ and $\beta_{13}$. The trilayer therefore supplies two design coordinates without requiring rotation of the support or repositioning of the nanoparticles. The bilayer and trilayer contain the same total α-$MoO_3$ thickness of 600 nm in Figs. 1–3. Only $\beta_{12}$ and $\beta_{13}$ are independent; the relative twist of layers 2 and 3 is $\beta_{23}=\beta_{13}-\beta_{12}$.

Interlayer electromagnetic coupling hybridizes the responses of the rotated flakes into collective polariton modes. In the trilayer, the additional internal twist angle provides a further degree of freedom for modifying their scattering strength, propagation direction, and coupling to the prescribed nanoparticle pair.

Figure 2(c) shows *P* over the trilayer internal-coordinate space ($\beta_{12}$,$\beta_{13}$) with $\beta_p$=0°. The maximum reaches 2.14×$10^{-14}$ W at (90°,0°), while the minimum is 1.66×$10^{-16}$ W, giving a max-to-min ratio of approximately 129. The corresponding common-threshold fractions and local angular-tolerance metrics are summarized in Table S1.

When both $\beta_p$ and $\beta_s$ are optimized, the bilayer reaches a maximum radiative power of 1.87×$10^{-14}$ W, whereas the trilayer reaches 2.14×$10^{-14}$ W, corresponding to a 14.4% higher maximum. When both the nanoparticle axis and the bottom-layer crystallographic frame are fixed at $\beta_p$ = 0°, the advantage becomes substantially larger: the trilayer maximum is 2.84 times the bilayer value, while its mean radiative power over all sampled internal twist configurations is 1.51 times the bilayer value. The larger max-to-min ratio of the bilayer reflects a greater switching contrast, whereas the trilayer provides a larger set of internal configurations supporting strong transfer under the fixed laboratory-frame geometry.

A ±10° variation of the bilayer twist angle $\beta_s$ retains, on average, 82.9% of its maximum radiative power. Separate ±10° variations of $\beta_{12}$ and $\beta_{13}$ retain 84.4% and 87.7% of the trilayer maximum, respectively, with $\beta_p$ fixed. These values characterize local one-angle tolerance rather than simultaneous uncertainty in both trilayer angles. Using common thresholds equal to 50% and 80% of the fixed-frame bilayer maximum, the fractions of sampled configurations are 36.67% and 15.56% for the bilayer and 51.75% and 31.85% for the trilayer, respectively (Table S1).

The momentum-resolved $\mathrm{Re}\left[G_{zz,\mathrm{sc}}\left(k_x, k_y\right)\right]$ maps in Figs. 2(*d*-f) reveal how the two internal angles reconstruct the collective polariton channels. The untwisted trilayer is weakly matched to the prescribed nanoparticle pair, whereas ($\beta_{12}$,$\beta_{13}$)=(90°,0°) produces strong receiver-directed coupling and the maximum of the trilayer map at $\beta_\mathrm{p}$=0°. At this configuration $I_\mathrm{sc}\left(\beta_\mathrm{p}\right)/I_0\left(\beta_\mathrm{p}\right)$≈309, compared with approximately 260 at the separately optimized bilayer map maximum. The normalized directionality metrics alone do not account for this difference; rather, the enhanced transfer reflects the combined effects of receiver alignment and scattering strength. This single-frequency diagnostic supports stronger receiver-directed trilayer scattering near resonance, while the complete frequency-integrated radiative power also includes direct coupling and vacuum–scattering interference.

Another pronounced, more visibly collimated response occurs at ($\beta_{12}$,$\beta_{13}$)=(140°,40°). Its nearly flat isofrequency-contour segments produce a canalized channel aligned with the *x*-axis and a radiative power of $1.76\times10^{-14}$ W. This value is below the maximum of the fixed-$\beta_p$ trilayer map at (90°,0°), showing that visual collimation or normalized directivity alone does not determine *P*; absolute modal weight and receiver matching must be considered together.

The three flakes jointly determine the hybridized polariton response and their contributions cannot be separated into additive components of the radiative power. The bottom flake defines the crystallographic reference frame, the top flake strongly influences the surface-field overlap because it is closest to the nanoparticles, and the middle flake couples to both outer layers and contributes to their modal hybridization. The relative importance of the middle and top layers therefore depends on the selected twist configuration.

Additional tests are provided in the Supporting Information. Figures S1-S3 examine reduced constituent-layer thickness, prescribed particle-axis directions, and thickness-controlled peak ordering; Table S1 summarizes the fixed-endpoint peak, mean radiative power over sampled internal twist configurations, common-target fractions, and equivalent one-angle tolerance. Equal constituent-layer thicknesses are used here to isolate the effect of the internal twist angles. Unequal thicknesses would introduce an additional design degree of freedom by modifying the hybridized polariton dispersion, surface-field overlap, and optimum twist configurations, but would not necessarily lead to a monotonic enhancement of the radiative heat transfer [66].

The corresponding $\mathrm{Re}\left[G_{zz,\mathrm{sc}}(k_x, k_y)\right]$ maps in Figs. S2(c,f,i) show receiver-directed polariton channels for ($\beta_p$,$\beta_{12}$,$\beta_{13}$)=(30°,92°,160°), (60°,124°,84°), and (90°,76°,106°). These are representative near-optimal grid configurations; for $\beta_p$=60°, the exact sampled-grid maximum occurs nearby at (126°,84°). The result therefore demonstrates a finite high-performance region rather than dependence on one uniquely specified angle pair.

Experimentally, the modeled direct-contact stack can be assembled by mechanically exfoliating the individual flakes and stacking them sequentially by dry transfer. The bottom flake first defines the laboratory frame, while the middle and top flakes are rotated to their target crystallographic orientations before being released into contact. The two internal twist angles are therefore independently selected during assembly or reconfiguration, rather than continuously adjusted after the interfaces have formed. Van der Waals adhesion, interfacial friction, and post-transfer relaxation may shift the final orientations, which should therefore be characterized after stacking together with the interface quality using optical methods [7, 67].

Figure S3 and Table S2 compare bilayer and trilayer structures at equal total thickness or equal constituent-layer thickness. The maximum radiative power depends on this choice. These two-coordinate maps scan ($\beta_p$,$\beta_s$) for the bilayer and ($\beta_{12}$,$\beta_{13}$) at $\beta_p = 0°$ for the trilayer, and are distinct from the fixed laboratory-frame angular comparison in Table S1. The 2×200 nm bilayer has the largest isolated maximum, $3.89\times10^{-14}$ W, compared with $1.87\times10^{-14}$ W for the 2×300 nm bilayer and $2.14\times10^{-14}$ W for the 3×200 nm trilayer. Peak ordering is therefore thickness dependent.

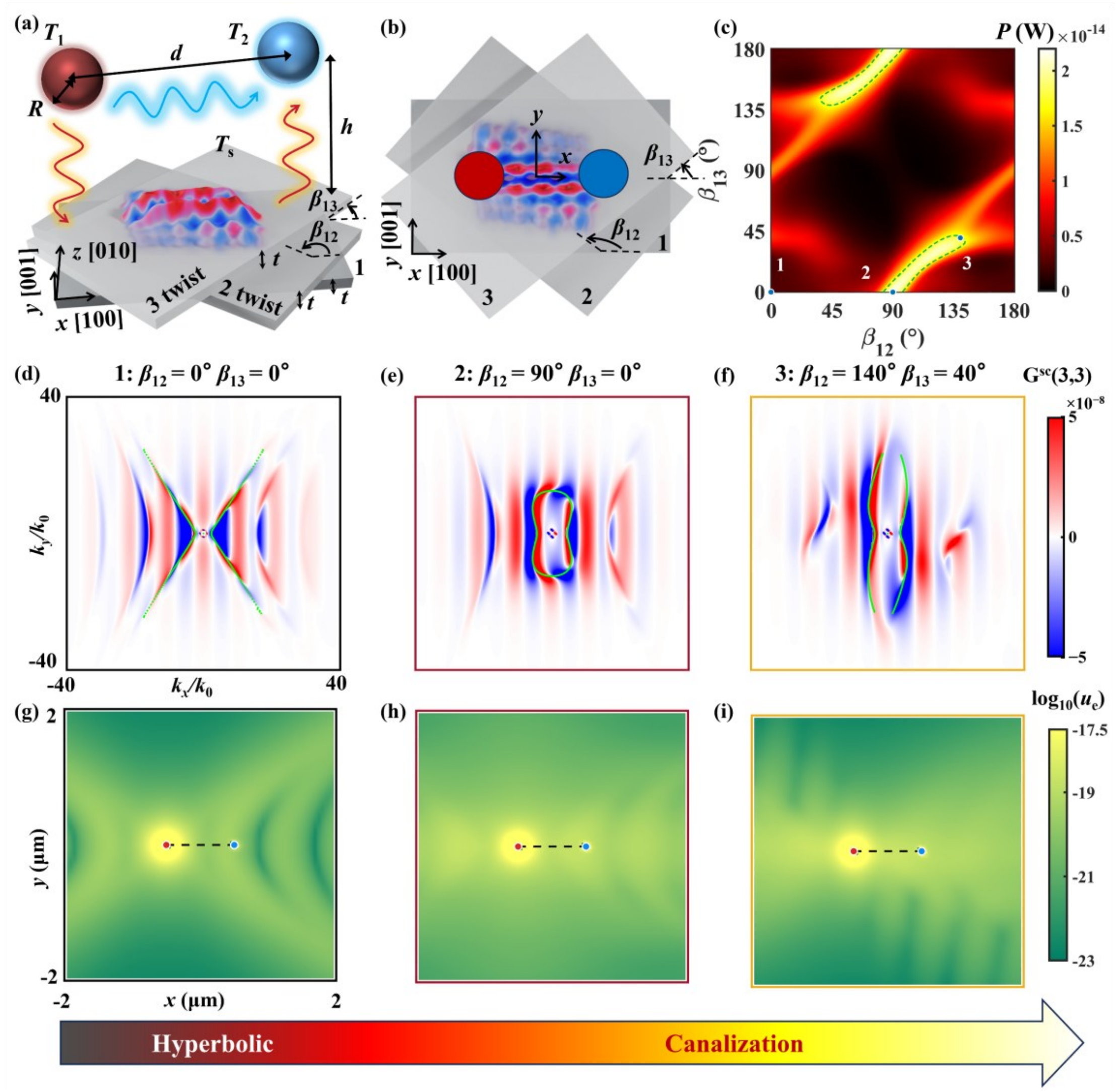

**Fig. 2.** NFRHT between two nanoparticles mediated by a twisted α-$MoO_3$ trilayer. (a) Schematic side view of two nanoparticles placed above a twisted α-$MoO_3$ trilayer. The three α-$MoO_3$ layers have the same thickness, $t$=200 nm. The bottom layer (layer 1) is kept fixed, while the middle and top layers (layers 2 and 3) are rotated relative to layer 1 by the interlayer twist angles $\beta_{12}$ and $\beta_{13}$, respectively. The two nanoparticles are aligned along the $x$-axis. (b) Top-view schematic of the twisted α-$MoO_3$ trilayer and nanoparticle arrangement. The layers are laterally infinite; the remaining parameters are the same as in Fig. 1. (c) Net radiative power $P$ between the two nanoparticles as a function of the two interlayer twist angles, $\beta_{12}$ and $\beta_{13}$. Numbered points identify the three configurations shown below, and green dashed outlines highlight high-power regions. (d–f) Momentum-space distributions of the scattering Green's function component $\mathrm{Re}\left[\mathbf{G}_{zz,sc}\left(k_x, k_y\right)\right]$ at 917 $cm^{-1}$ for representative twist-angle combinations: (d) $\beta_{12}$=0°, $\beta_{13}$=0°; (e) $\beta_{12}$=90°, $\beta_{13}$=0°; and (f) $\beta_{12}$=140°, $\beta_{13}$=40°. The green curves represent the isofrequency contours of the twisted α-$MoO_3$ trilayer. (g–i) Corresponding emitter-generated spectral electric-field energy-density maps at 917 $cm^{-1}$ in a plane 150 nm above the upper surface. Red and blue circles mark the emitter and receiver, while black dashed lines indicate their connecting direction. The bottom arrow schematically indicates the evolution from hyperbolic polariton propagation to canalized phonon-polariton transport.

After establishing the angular response, we examine how the result changes with the selected nanoparticle resonance. At each sampled target resonance $\Omega$, the Drude plasma parameter is set according to $\omega_p \simeq \sqrt{3}\Omega$, corresponding to the weak-damping quasistatic resonance condition for a sphere. Figure 3 compares a family of particle designs whose dipolar resonances are centered at different target values, rather than the spectrum of one fixed particle. The three selected trilayer configurations retain pronounced enhancement over the sampled resonance range, showing that the behavior is not limited to the map maximum, although its precise range and magnitude remain configuration dependent.

Figure 3(a) compares three trilayer configurations selected to sample distinct parts of the broad high-$P$ region in Fig. 2(c). The gold, purple, and green curves correspond to radiative powers of $1.30\times10^{-14}$, $1.76\times10^{-14}$, and $2.14\times10^{-14}$ W, respectively, for particles resonant at 917 $cm^{-1}$. They represent moderate-, high-, and maximum-power states rather than cases selected to maximize the resonance range.

To understand this resonance-tuning behavior, we analyze the momentum-space distribution of $\mathrm{Im}(r_{pp})$, as shown in Figs. 3(b-j). The rows correspond to 930, 910, and 890 $cm^{-1}$, while the columns compare the slab configurations indicated above the panels. For the twisted α-$MoO_3$ bilayer with $\beta_p$=0° and $\beta_s$=90°, extended canalization-like features occur at $\Omega$=930 $cm^{-1}$, but they weaken or reshape at $\Omega$=910 and 890 $cm^{-1}$, making the transfer sensitive to the selected particle resonance. By contrast, the trilayer's two internal twist coordinates provide more flexible dispersion engineering. Its channels remain pronounced at 930 $cm^{-1}$, become more closed but strongly anisotropic at 910 $cm^{-1}$, and retain finite directional features at 890 $cm^{-1}$. The selected trilayers therefore maintain stronger receiver-directed polaritonic coupling over the sampled family of nanoparticle resonances. The contour shapes and degree of collimation remain frequency dependent.

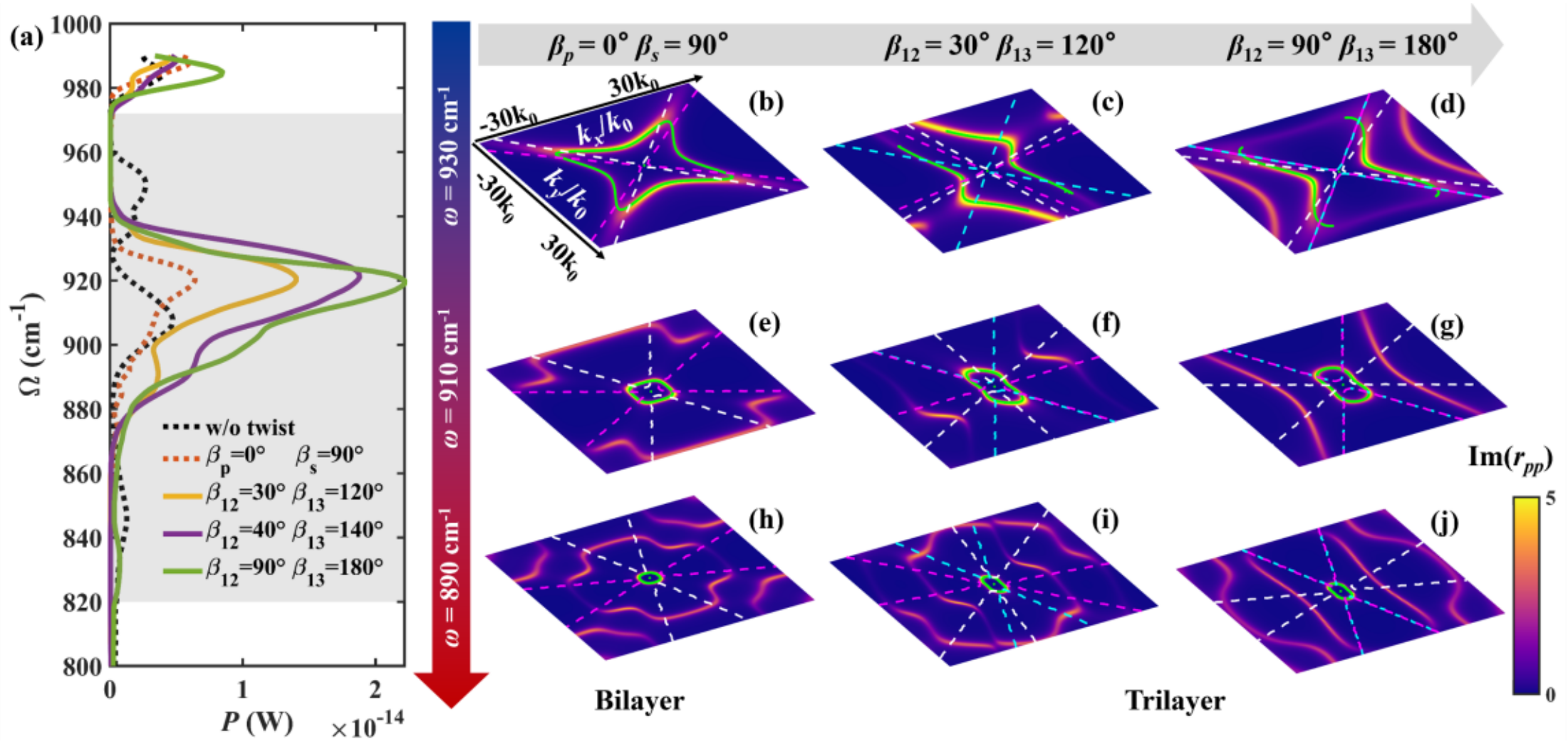

**Fig. 3.** Spectral robustness of NFRHT mediated by twisted α-$MoO_3$ trilayers. (a) Net radiative power $P$ between the two nanoparticles as a function of the nanoparticle resonance frequency $\Omega$. At each $\Omega$, the nanoparticle plasma parameter is adjusted to place its dipolar resonance at $\Omega$; the curves therefore compare a family of particle designs. The dashed curves correspond to the reference configurations without twist and to the twisted bilayer with $\beta_p$=0° and $\beta_s$=90°. The solid curves correspond to representative α-$MoO_3$ trilayers with $\beta_{12}$=30°, $\beta_{13}$=120°; $\beta_{12}$=40°, $\beta_{13}$=140°; and $\beta_{12}$=90°, $\beta_{13}$=180°. The shaded region denotes the Reststrahlen band of α-$MoO_3$. (b–j) Momentum-space distributions of the imaginary part of the reflection coefficient, Im($r_{pp}$), at different frequencies and twist-angle combinations. The rows correspond to $\Omega$=930, 910, and 890 cm$^{-1}$, respectively. The left column shows the twisted bilayer with $\beta_p$=0° and $\beta_s$=90°, while the middle and right columns show twisted trilayers with $\beta_{12}$=30°, $\beta_{13}$=120° and $\beta_{12}$=90°, $\beta_{13}$=180°, respectively. The green curves indicate the corresponding isofrequency contours. The cyan, white, and magenta dashed curves represent the isofrequency contours of the individual α-$MoO_3$ layers.

Finally, Fig. 4(a) compares the distance dependence of $P/P_{\mathrm{BB}}$ for vacuum, an untwisted trilayer, a single α-$MoO_3$ slab, a twisted bilayer, and several twisted trilayers. The far-field reference for two identical spheres is $P_{\mathrm{BB}}(d) = \sigma A F_{12}(d)[{T_1}^4 - {T_2}^4]$, with $A = 4\pi R^2$ and $F_{12}(d) = \left[1-\sqrt{1-(R/d)^2}\right]/2$; $d$ is the center-to-center separation. For d≫R, $P_{\mathrm{BB}} \propto d^{-2}$, whereas direct quasistatic dipole exchange obeys $P \propto d^{-6}$, giving $P/P_{\mathrm{BB}} \propto d^{-4}$. For this representative distance comparison, every constituent slab is 200 nm thick; the single-slab, bilayer, and trilayer total thicknesses are therefore 200, 400, and 600 nm. Figure 4 is not a matched-material-volume comparison, unlike the fixed-600-nm comparison in Figs. 1–3.

As $d$ increases, the slab-mediated polaritonic contribution becomes dominant. Without α-$MoO_3$, $P$ approaches the blackbody scale rapidly. The single α-$MoO_3$ slab and twisted bilayer sustain directional surface-wave channels, while selected trilayers maintain the largest long-range response for the configurations shown. Because the total material thickness differs in this figure, these curves demonstrate representative propagation behavior rather than isolating layer number alone.

The selected trilayer responses remain several orders of magnitude above the α-$MoO_3$-free reference across a broad distance interval. This persistence reflects collective twist-reconstructed polariton branches whose direction, modal weight, attenuation length, and surface overlap favor the prescribed particle positions. Adding further layers would not necessarily increase NFRHT monotonically. Additional interfaces and twist coordinates can modify the collective polariton spectrum, but their contribution to the nanoparticle–nanoparticle transfer depends on directional matching, damping, modal weight, and surface-field overlap.

Figure 4(b) compares the frequency-integrated Green-tensor intensities of the vacuum and scattering contributions:

$$\mathcal{G}_{\mu}(d) = \int_{0}^{\infty} \mathrm{Tr}\left[G_{\mu}(d,\omega) {G_{\mu}}^{\dagger}(d,\omega)\right] \mathrm{d}\omega$$

Here $\mu \in \{0,\mathrm{sc}\}$ labels the vacuum and scattering contributions. The trace sums the squared magnitudes of all nine Cartesian tensor components. Since $G$ has units $\mathrm{m}^{-1}$ and the integration is over angular frequency, $\mathcal{G}_{\mu}$ has units $\mathrm{m}^{-2}\mathrm{s}^{-1}$; no Planck or particle-polarizability weight is included. In the quasistatic regime, $G_0$ scales in amplitude approximately as $d^{-3}$, so the plotted Green tensor intensity follows an approximate $d^{-6}$ dependence. The scattering contribution decays more slowly over the subwavelength distance range considered and eventually exceeds the vacuum contribution, indicating that slab-mediated polaritonic propagation dominates the electromagnetic coupling between the two particle positions in this regime.

The oscillations of the scattering Green-tensor intensity arise from coherent interference among polaritonic contributions with different in-plane wavevectors. Their relative propagation phases vary with lateral separation, producing alternating constructive and destructive interference. Internal reflections within the finite-thickness stack are already contained in the multilayer reflection matrix and determine the dispersion and modal weights of these contributions, while twisting modifies their wavevectors and relative amplitudes. The complete radiative power additionally contains vacuum–scattering interference, so its extrema need not coincide with those of the scattering-only Green-tensor intensity.

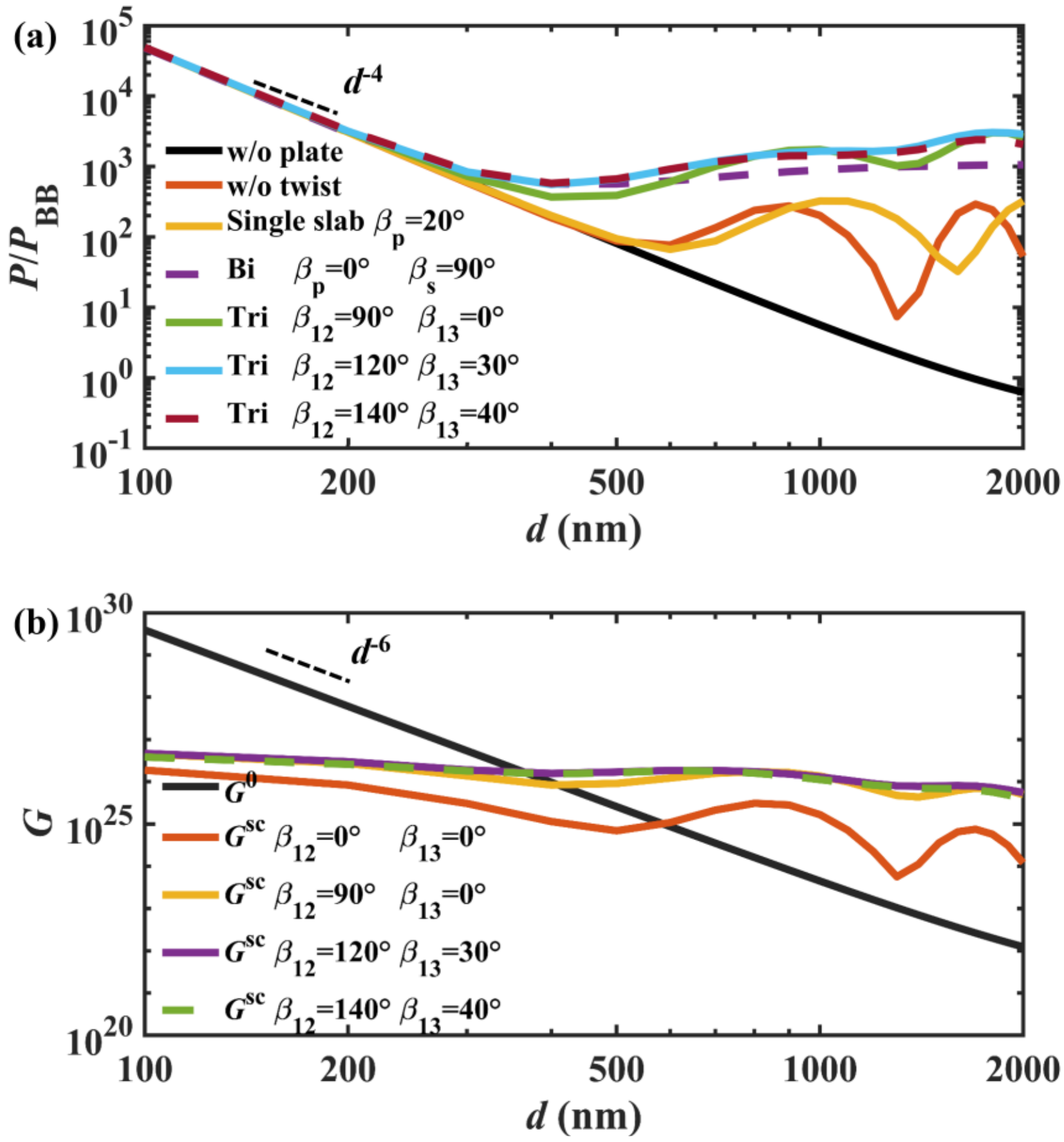

**Fig. 4.** Distance dependence of NFRHT between nanoparticles mediated by α-$MoO_3$ polaritonic layers. (a) Normalized net radiative power $P/P_{BB}$, as a function of the interparticle distance $d$ for the reference case without α-$MoO_3$, an untwisted α-$MoO_3$ trilayer, a single 200-nm α-$MoO_3$ slab (labeled "Single slab") with $\beta_p$=20°, a twisted α-$MoO_3$ bilayer with $\beta_p$=0° and $\beta_s$=90°, and twisted α-$MoO_3$ trilayers with representative interlayer twist-angle combinations. Every constituent slab in this panel is 200 nm thick, so total thickness is not matched across layer numbers. The black dashed line indicates an approximate $d^{-4}$ scaling at short distances. (b) Frequency-integrated Green-tensor intensities of the vacuum and scattering contributions; all nine tensor components are included. The dashed guide indicates the approximate $d^{-6}$ scaling of $G^0$.

In conclusion, we have used the established twisted α-$MoO_3$ trilayer platform to examine radiative heat transfer between a prescribed nanoparticle pair. When both the nanoparticle axis and the bottom-layer crystallographic frame are fixed, the trilayer maximum and mean radiative powers are 2.84 and 1.51 times the corresponding bilayer values at equal total thickness. The second internal twist coordinate enlarges the set of configurations supporting strong transfer, and separate ±10° perturbations of either internal angle retain more than 80% of the corresponding maximum. When the bilayer is also allowed to optimize its global orientation, the trilayer maximum remains 14.4% higher for the parameters considered, while the bilayer provides the larger switching contrast. Green-tensor diagnostics and real-space maps associate strong transfer with both sufficient scattering strength and favorable receiver alignment. The selected trilayer configurations also sustain enhanced coupling over the sampled family of particle resonances and at long distances. These results establish the benefit of internal twist engineering for the geometries considered, without implying a universal increase with layer number or robustness against all material and geometric variations.

## Supporting Information

Fluctuational-electrodynamic model and dipole-approximation validity; numerical implementation and convergence; α-$MoO_3$ Lorentz dielectric parameters; thickness and nanoparticle-orientation tests; fixed-endpoint peak, mean radiative power over sampled internal twist configurations, common-target fractions, and one-angle tolerance; and canalization diagnostics.

## Data Availability

Data will be made available upon reasonable request.

## Note

The authors declare no competing financial interest.

## Acknowledgments

P. A-G acknowledges the Spanish Ministry of Science and Innovation under the National Plan grant PID2022-141304NB-I00, the European Research Council under Consolidator grant no. 101044461, TWISTOPTICS, and Agencia SEKUENS (Asturias) under grant UONANO IDE/2024/000678 with the support of FEDER funds.